\documentclass[onecolumn,superscriptaddress,preprintnumbers,amsmath,aps]{revtex4}

\usepackage{amsmath}
\usepackage[pdftex]{graphicx}
\usepackage{subfigure}

\usepackage{color}
\usepackage{dcolumn}
\usepackage{amsfonts}
\usepackage{bm}
\usepackage{amssymb}

\begin{document}
	
\title{Single-pixel 3D imaging with time-based depth resolution}

\author{Ming-Jie~Sun}
\email{mingjie.sun@buaa.edu.cn}
\affiliation{Department of Opto-electronic Engineering, Beihang University, Beijing, 100191, China}
\affiliation{SUPA, School of Physics and Astronomy, University of Glasgow, Glasgow, G12 8QQ, UK}

\author{Matthew.~P.~Edgar}
\affiliation{SUPA, School of Physics and Astronomy, University of Glasgow, Glasgow, G12 8QQ, UK}

\author{Graham~M.~Gibson}
\affiliation{SUPA, School of Physics and Astronomy, University of Glasgow, Glasgow, G12 8QQ, UK}

\author{Baoqing~Sun}
\affiliation{SUPA, School of Physics and Astronomy, University of Glasgow, Glasgow, G12 8QQ, UK}

\author{Neal~Radwell}
\affiliation{SUPA, School of Physics and Astronomy, University of Glasgow, Glasgow, G12 8QQ, UK}

\author{Robert~Lamb}
\affiliation{Selex ES, Edinburgh, UK}

\author{Miles~J.~Padgett}
\email{miles.padgett@glasgow.ac.uk}
\affiliation{SUPA, School of Physics and Astronomy, University of Glasgow, Glasgow, G12 8QQ, UK}

\begin{abstract}
{\bf Time-of-flight three dimensional imaging is an important tool for many applications, such as object recognition and remote sensing. Unlike conventional imaging approach using pixelated detector array, single-pixel imaging based on projected patterns, such as Hadamard patterns, utilises an alternative strategy to acquire information with sampling basis. Here we show a modified single-pixel camera using a pulsed illumination source and a high-speed photodiode, capable of reconstructing 128$\times$128 pixel resolution 3D scenes to an accuracy of $\sim$ 3\,mm at a range of $\sim$ 5\,m. Furthermore, we demonstrate continuous real-time 3D video with a frame-rate up to 12\,Hz. The simplicity of the system hardware could enable low-cost 3D imaging devices for precision ranging at wavelengths beyond the visible spectrum.}
\end{abstract}

\maketitle

\section*{Introduction}

Whilst a variety of 3D imaging technologies are suited for different applications, time-of-flight (TOF) systems have set the benchmark for performance with regards to a combination of accuracy and operating range. Time-of-flight imaging is performed by illuminating a scene with a pulsed light source and observing the back-scattered light. Correlating the detection time of the back-scattered light with the time of the illumination pulse allows the distance, $d$, to objects within the scene to be estimated by $d=tc/2$, where $t$ is the TOF and $c$ is the propagation speed of light. 

The transverse spatial resolution of the image obtained is retrieved either by using a pixelated array or by using a single-pixel detector with a scanning approach \cite{Schwarz2010,McCarthy2009,Shapiro2008,Bromberg2009,McCarthy2013,Radwell2014,Edgar2015}. In both cases, the inherent speed of light demands the use of detectors with a fast response time and high-speed electronic readout in order to obtain high precision depth resolution. Advances in sensor development have enabled the first TOF single-photon avalanche detector (SPAD) array cameras to enter the commercial, having a resolution of $32\times32$ pixels, however such devices are still in their infancy \cite{niclass2005,richardson2009,entwistle2012}. On contrary, there are mature single-pixel detectors in the market which provide stable time-resolved measurements, and by employing compressed sensing principles for image reconstruction, which takes advantage of the sparsity in natural scenes, the acquisition times of the scanning approach is largely reduced \cite{Baraniuk2007,Duarte2008,Mei2011Real,Hatef2013Deterministic,Herman2013A}.

Recently there have been some interesting developments in 3D imaging utilising single-pixel detectors. One technique utilises structured illumination and spatially separated photodiodes to obtain multiple images with different shading properties from which 3D images can be reconstructed via photometric stereo \cite{Sun2013}. Another scheme scans a scene, pixel by pixel, using a pulsed illumination source and measures the reflected light using an avalanche photodiode (APD), whereupon the first detected photon is used to recover depth and reflectivity via TOF \cite{Kirmani2014}. An alternative method for scanning a scene and recovering depth and reflectivity via TOF has also been demonstrated utilising structured pulsed illumination \cite{kirmani2011exploiting,Howland2011,Colaco2012,Howland2013}.

Among the mentioned demonstrations, many \cite{McCarthy2013, Kirmani2014,Howland2011,Colaco2012,Howland2013} employed photon counting detection (i.e. Geiger-mode), which is well suited for low light level imaging. However, one limitation of photon counting detectors is the inherent electronic dead-time between successive measurements, often 10's of nanoseconds, which prohibits the retrieval of short range timing information from a single illumination pulse. Instead, an accurate temporal response from a 3D scene requires summing the data over many back-scattered photons and hence many illumination pulses (usually several hundreds or thousands \cite{Colaco2012,Howland2013}). In contrast, as first demonstrated by Kirmani \cite{kirmani2011exploiting}, a high-speed photodiode can retrieve the temporal response from a single illumination pulse, which can be advantageous in certain circumstances, for instance when the reflected light intensity is comparatively large. Incidentally, photon counting cannot operate under such conditions since the detection will always be triggered by back-scattered photons from the nearest, or most reflective, object, rendering more distant objects invisible.

In this paper, we present a single-pixel 3D imaging system employing pulsed structured illumination and a high-speed (short response time) photodiode for sampling the time-varying intensity of the back-scattered light from a scene. We show that by using an analogue photodiode to record the full temporal form of the back-scattered light, along with our original 3D reconstruction algorithm, it is possible to recover surface profiles of objects with an accuracy much better than that implied by the finite temporal bandwidth of the detector and digitisation electronics. At distances of $\sim5\,{\rm m}$ we demonstrate a range profile accuracy of $\sim3\,{\rm mm}$ with image resolutions of $128\times128$ pixels, whilst simultaneously recovering reflectivity information of the object. This accuracy is achieved despite a detection bandwidth and a digitisation interval corresponding to distances of $150\,{\rm mm}$ and $60\,{\rm mm}$ respectively. We further demonstrate that by employing a compressive sampling scheme, the system is capable of performing continuous real-time 3D video with a frame-rate up to $12\,{\rm Hz}$.

\begin{figure*}[h]
	\centering
	\includegraphics[width=0.9\textwidth]{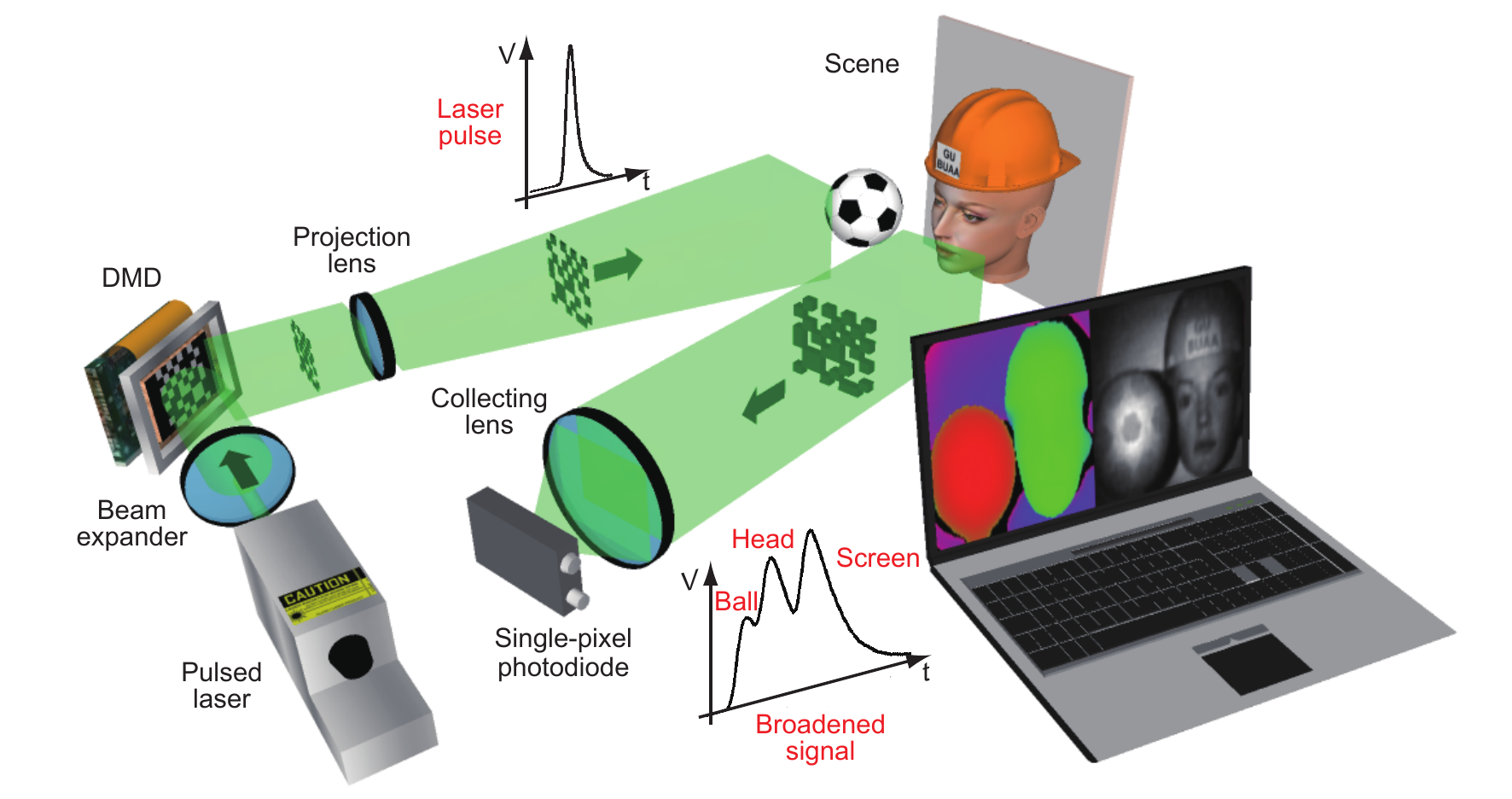}
	\caption{\textbf{Single-pixel 3D imaging system.} A pulsed laser uniformly illuminates a digital-micromirror-device (DMD), used to provide structured illumination onto a scene, and the back-scattered light is collected onto a photodiode. The measured light intensities are used in a 3D reconstruction algorithm to reconstruct both depth and reflectivity images.}
	\label{Figure1}
\end{figure*}

\section*{Experimental setup}
The single-pixel 3D imaging system, illustrated in Fig.\,\ref{Figure1}, consists of a pulsed laser and a digital-micro mirror-device (DMD) to provide time varying structured illumination. A high-speed photodiode is used in conjunction with a fresnel lens condenser system to measure the back-scattered intensity resulting from each pattern. The analogue photodiode output is passed through a low-noise amplifier and sampled using a high-speed digitiser. In our work we chose to use the Hadamard matrices \cite{pratt1969} for providing structured illumination. In order to remove sources of noise, such as fluctuations in ambient light levels, we obtain differential signals by displaying each Hadamard pattern, followed by its inverse pattern, and taking the difference in the measured intensities \cite{sun2013differential,smj2015PNGI}. Detailed information about experimental setup is provided in Methods.

\begin{figure*}[h]
	\centering
	\includegraphics[width=0.9\textwidth]{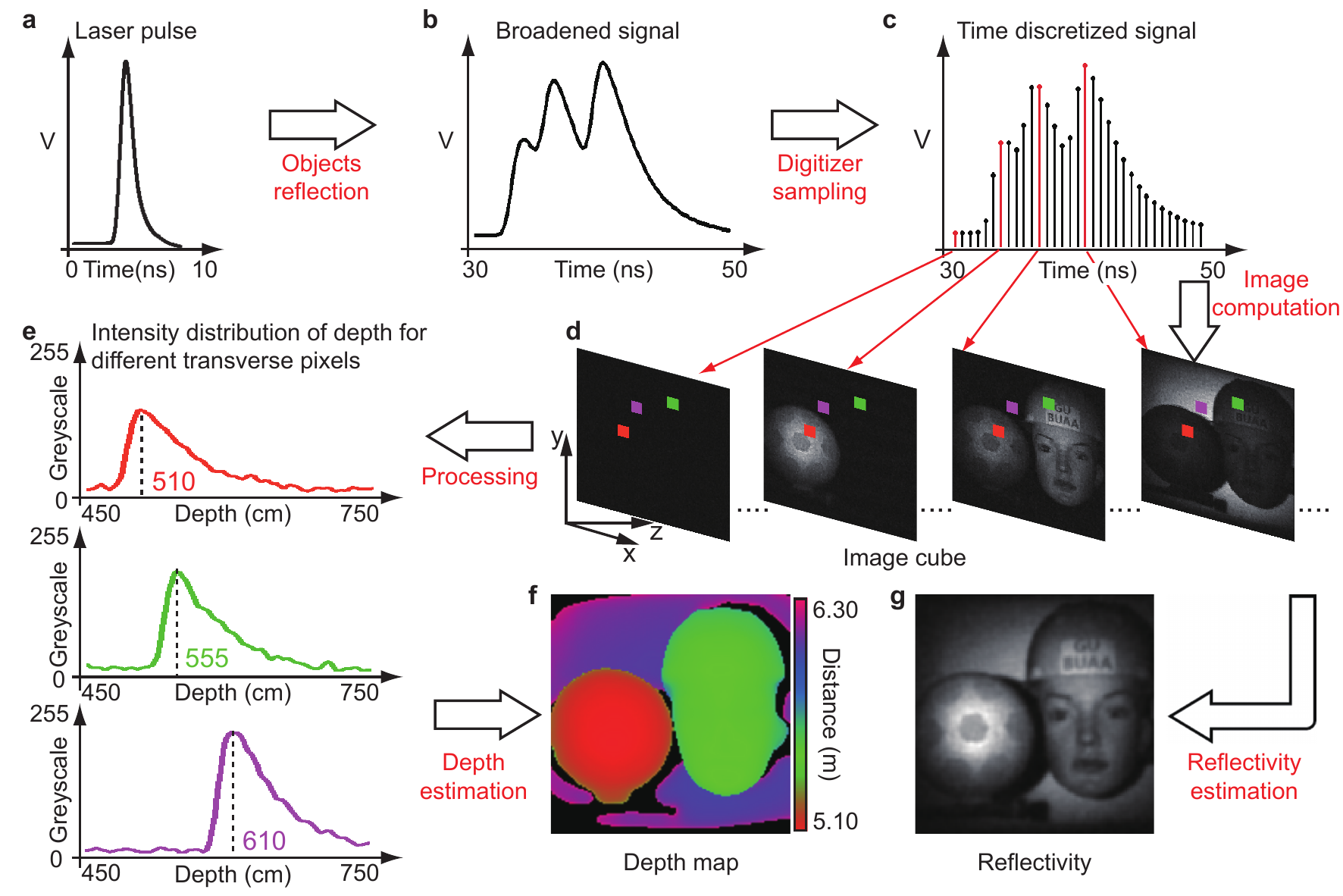}
	\caption{\textbf{Overview of the reconstruction algorithm.} The incident laser pulses (a) back-scattered from a 3D scene are temporally broadened (b) and discretely sampled using a high-speed digitizer (c). An image cube (d) is obtained using a reconstruction algorithm, from which the depth map (f) and the reflectivity (g) of the scene can be estimated.}
	\label{Figure2}
\end{figure*}

A 3D image of the scene is reconstructed utilising the time-varying back-scattered intensities (measured for each output pulse of the laser) and the associated set of $N$ patterns used to provide the structured illumination. An overview of the reconstruction algorithm is shown in Fig.\,\ref{Figure2} (the result in this diagram represents a scene of three objects $\sim0.5\,{\rm m}$ apart in depth). The high-speed digitiser converts the amplified analog signals (Fig.\,\ref{Figure2}b) into discrete data points (Fig.\,\ref{Figure2}c), which are subsequently processed by the computer algorithm. Whereas typical single-pixel imaging schemes use the integrated signal for each illumination pattern to reconstruct a 2D image, our algorithm utilises $M$ discretely sampled intensity points from the time varying signal to reconstruct $M$ 2D images, resulting in an $\rm x,~y,~z$ image cube (Fig.\,\ref{Figure2}d). In the image cube, each transverse pixel $\rm (x, y)$ has an intensity distribution (Fig.\,\ref{Figure2}e) along the longitudinal axis $\rm (z)$, which is related to the temporal shape of the pulse, the detector response, the readout digitisation and the pixel depth and reflectivity information.

To enhance the range precision beyond the limits imposed by the sampling rate of the system, methods such as parametric deconvolution \cite{kirmani2011exploiting,Colaco2012} and curve fitting can be employed. However, often these methods can be computationally intensive, which makes them unsuitable for real-time applications. Instead we choose to apply cubic spline interpolation to the reconstructed temporal signal at each pixel location, which introduces minimal computational overhead. The depth of the scene (Fig.\,\ref{Figure2}f) is subsequently determined by finding the maximum in these interpolated signals. In addition, the scene reflectivity (Fig.\,\ref{Figure2}g) can be calculated by averaging the image cube along the longitudinal axis. Utilizing both the depth and reflectivity information, a 3D image of the scene is then reconstructed. 

It is worth mentioning that, with the assumption that there is only one surface at each transverse pixel location, our depth estimation (see detail in Methods) works well for scenes that have smooth features, such as the mannequin head and ball, since the reconstructed temporal signals should be slowly varying between sample points. We note that the depth accuracy is limited by the amplitude noise on the data points, over-interpolation only increases the depth precision, but not necessarily the accuracy. In addition, more interpolation adds more processing time, therefore in our experiments we chose to interpolate by a factor of 5 times when investigating the static scenes with 20 pulses per pattern (see Fig.\,\ref{Figure3}-\ref{Figure5}), and 4 times when investigating scenes with motion (see Fig.\,\ref{Figure6}), balancing the computational overhead and 3D image quality.

\section*{Results}

\begin{figure*}[h]
	\centering
	\includegraphics[width=0.9\textwidth]{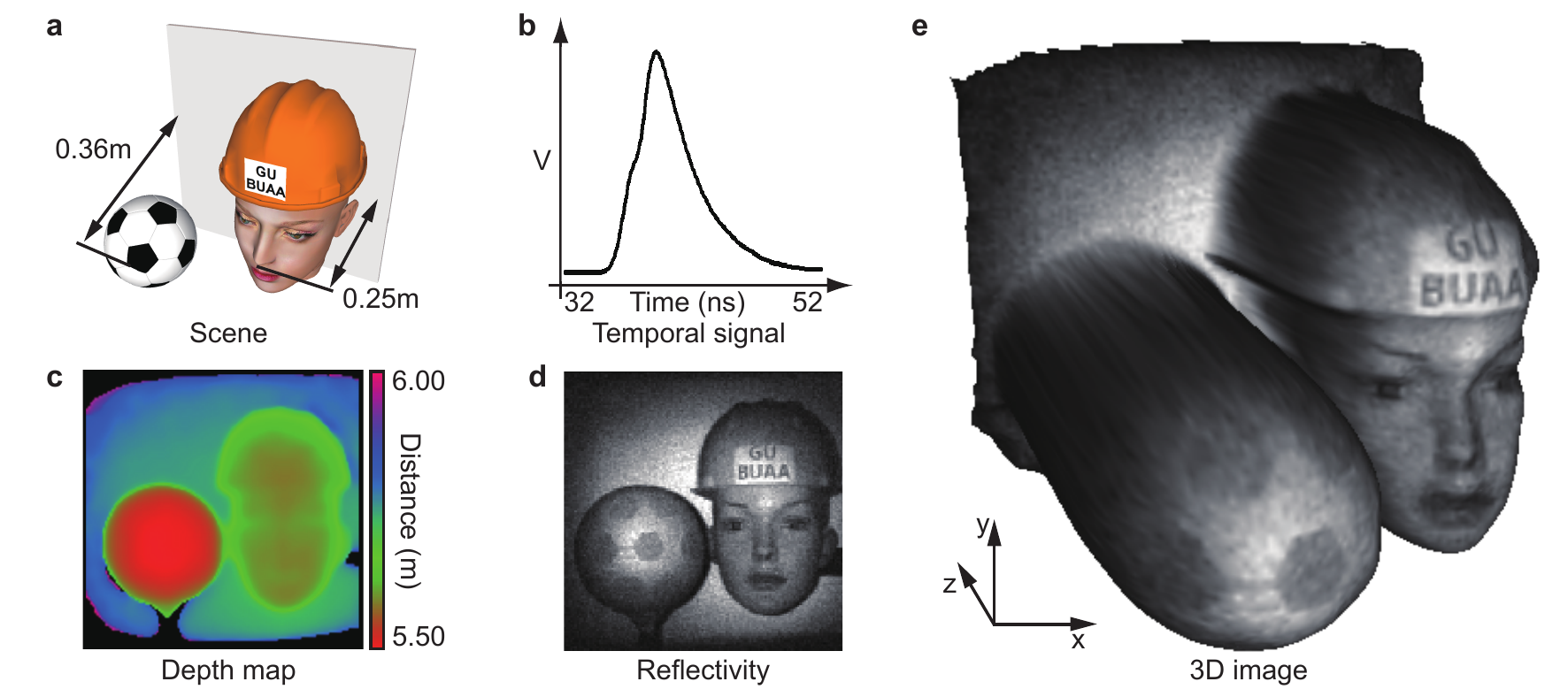}
	\caption{\textbf {3D image of a scene.} (a) Illustration of the scene containing multiple objects in close proximity. (b) Reflected intensity measured for uniform illumination, indicating temporally indistinguishable objects. (c) The estimated depth map of the scene. (d) The reconstructed scene reflectivity. (e) A textured 3D image of the scene.}
	\label{Figure3}
\end{figure*}

In one experiment, a scene containing a $140\,{\rm mm}$ diameter polystyrene soccer ball, a life size skin-tone mannequin head and a screen was located at a distance of $\sim 5.5\,{\rm m}$ from the imaging system (Fig.\,\ref{Figure3}a). The objects were closely separated in distance such that the total depth of the scene was $\sim 360\,{\rm mm}$. A complete Hadamard set of 16,384 patterns, and their inverse, were used as the structured illumination and the back-scattered intensities measured for reconstructing a $128 \times 128$ pixel resolution 3D image (Fig.\,\ref{Figure3}e). The illumination time of each pattern was $2.66\,{\rm ms}$, corresponding to twenty laser pulses. The total time for acquisition, data transfer from the digitizer buffer to the computer, and image processing was $\sim 130$ seconds. The 3D reconstruction shown in Fig.\,\ref{Figure3}e exhibits distinguishable features, such as the profile of the head and the ball. 

\begin{figure*}[t]
	\centering
	\includegraphics[width=0.9\textwidth]{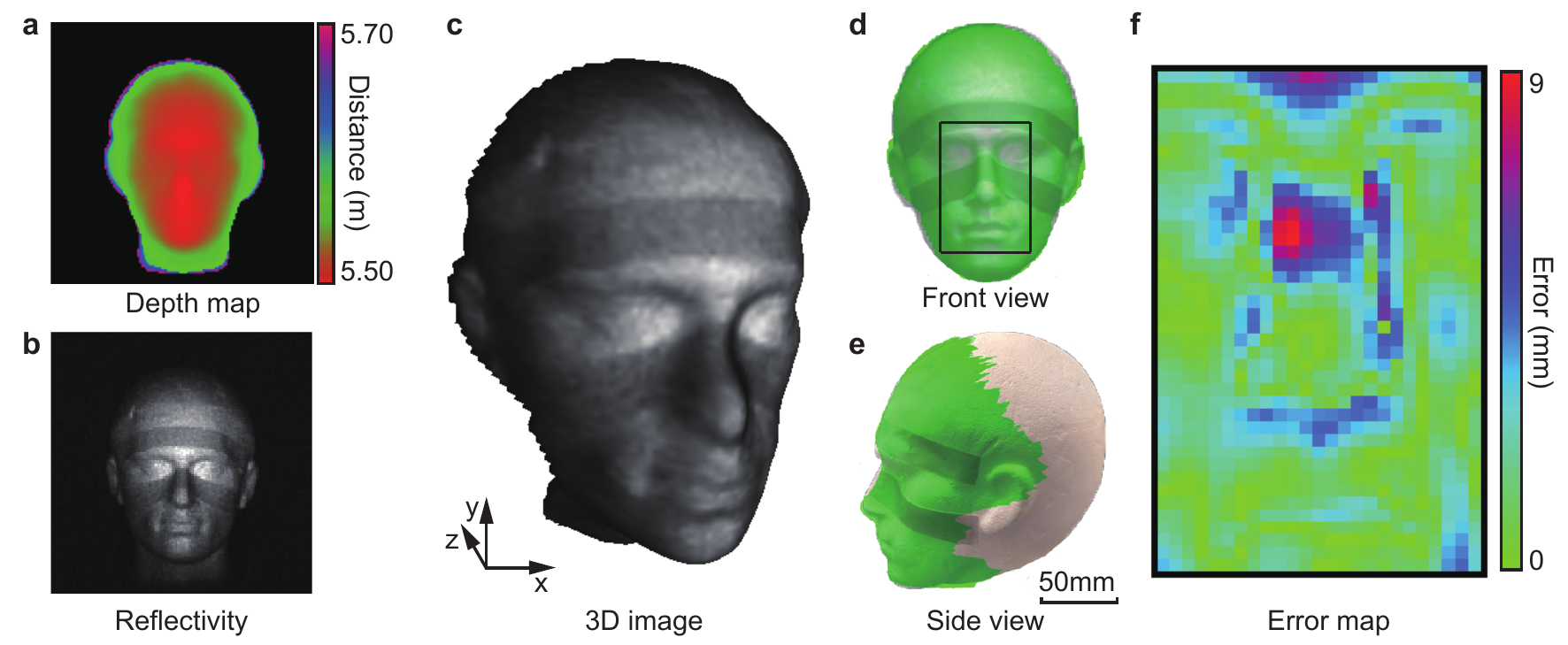}
	\caption{\textbf{Quantitative analysis of 3D reconstruction.} The depth estimation (a), reflectivity (b) and 3D reconstruction (c) of a white polystyrene mannequin head at a range of $\sim5.5\,{\rm m}$. Superposed depth reconstruction and photograph of the mannequin head, viewed from the front (d) and side (e). (f) For a chosen region of interest an error map showing the absolute differences between our depth result and that obtained using a stereophotogrammetric camera system.}
	\label{Figure4}
\end{figure*}

In order to quantitatively determine the accuracy of our 3D imaging system, the scene was modified to contain only a polystyrene mannequin head ($180\times270\times250\,{\rm mm}$), for which we had reference 3D data obtained via a high-accuracy stereophotogrammetric camera system \cite{Khambay2008,Sun2013}. The mannequin head was located at a distance of $5.5\,{\rm m}$ from the imaging system. To further demonstrate the system capability for retrieving reflectivity in addition to the depth, two grey stripes were placed on the head. Performing the same acquisition and imaging processing used in Fig.\,\ref{Figure3}, we obtained the results shown in Fig.\,\ref{Figure4}a-c. Figure \ref{Figure4}d\,\&\,e show the front and side view comparisons between our 3D reconstruction (green) and photographs of the head (white), respectively. After lateral and angular registration and subsequent depth scaling, an error map representing the absolute differences for a chosen region of interest was obtained (shown in Fig.\,\ref{Figure4}f). From this comparison we find our single-pixel 3D imaging system has a root mean square error (RMSE) of $2.62\,{\rm mm}$.

\begin{figure*}[h]
	\centering
	\includegraphics[width=0.9\textwidth]{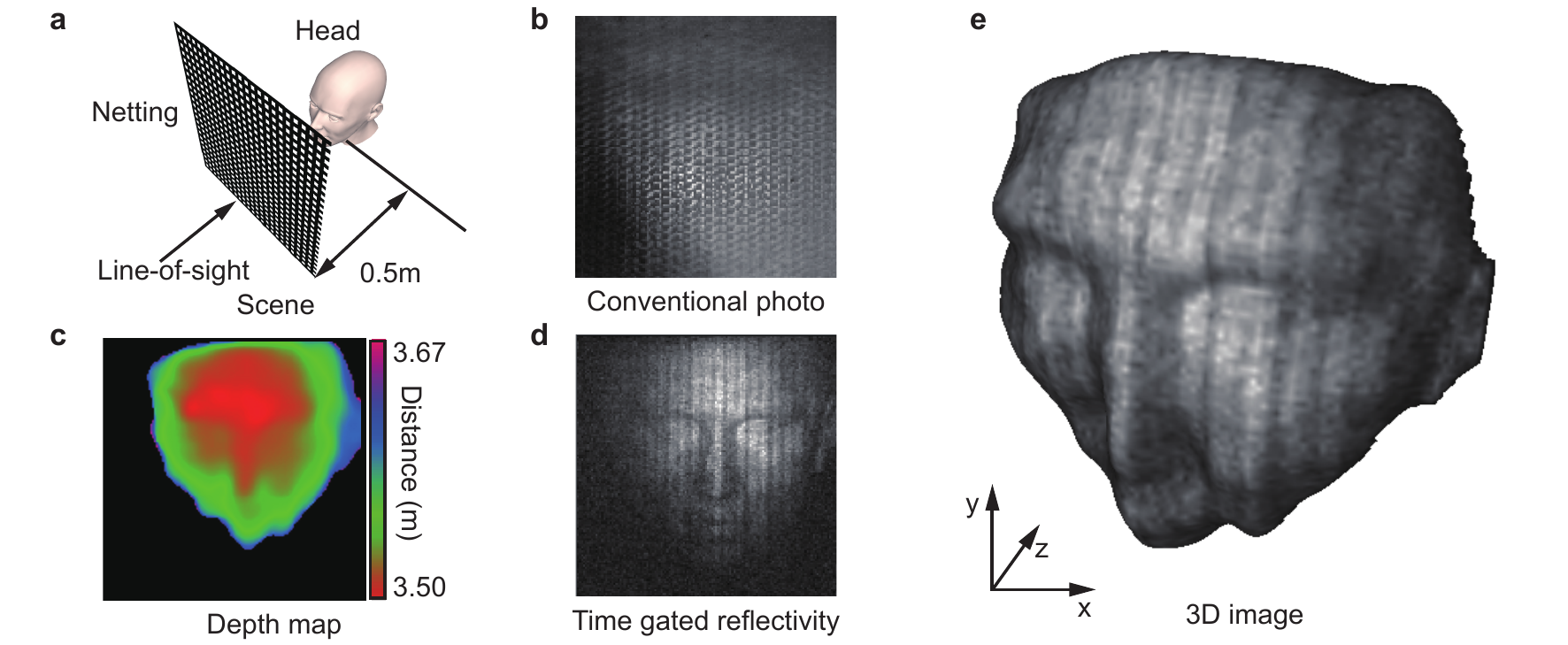}
	\caption{\textbf{3D imaging through obscuring material.} (a) Illustration of scene containing a mannequin head with black netting material obscuring the line-of-sight. (b) A photograph of the scene taken from the perspective of the 3D imaging system. The scene depth (c) and reflectivity (d) reconstructed by time-gating the measured intensity data. (e) 3D reconstruction of the mannequin head.}
	\label{Figure5}
\end{figure*}

One advantage of time-resolved imaging is the ability to distinguish objects at different depths, by artificially time-gating the measured intensity. In certain cases this enables obscuring objects to be isolated from objects of interest. Similar to previous demonstrations \cite{Howland2011}, we constructed a 3D scene containing a polystyrene mannequin head (located at a distance of $\sim 3.5\,{\rm m}$) and black-coloured netting used to obstruct the line-of-sight (located at a distance of $\sim 3\,{\rm m}$), as illustrated in Fig.\,\ref{Figure5}a. An image of the scene taken using a conventional camera is shown in Fig.\,\ref{Figure5}b, which contains no information of the head because of the black netting's obscuring. Performing the same acquisition and imaging processing used in Fig.\,\ref{Figure3} and \ref{Figure4}, along with an artificial gating on the photodiode data to ensure no reflected signals from the black netting are included in the 3D reconstruction, we obtained the results shown in Fig.\,\ref{Figure5}c-e. As before, we note the characteristic features of the mannequin head can be resolved.

In addition to obtaining high-quality 3D images of static scenes, many applications demand video frame-rates for motion tracking in dynamic scenes. A key merit of single-pixel imaging is the ability to take advantage of the sparsity of the scene and use compressive sensing to reduce the acquisition time. Most compressive sensing schemes are performed by minimizing a certain measure of the sparsity, such as $L_1$-norm, to find the sparsest image as the optimal reconstruction of the scene. However, for resolutions greater than $32 \times 32$ pixels, the time taken by the construction algorithm often prohibits real-time application \cite{Howland2013, abmann2013compressive}. 

\begin{figure*}[t]
	\centering
	\includegraphics[width=0.9\textwidth]{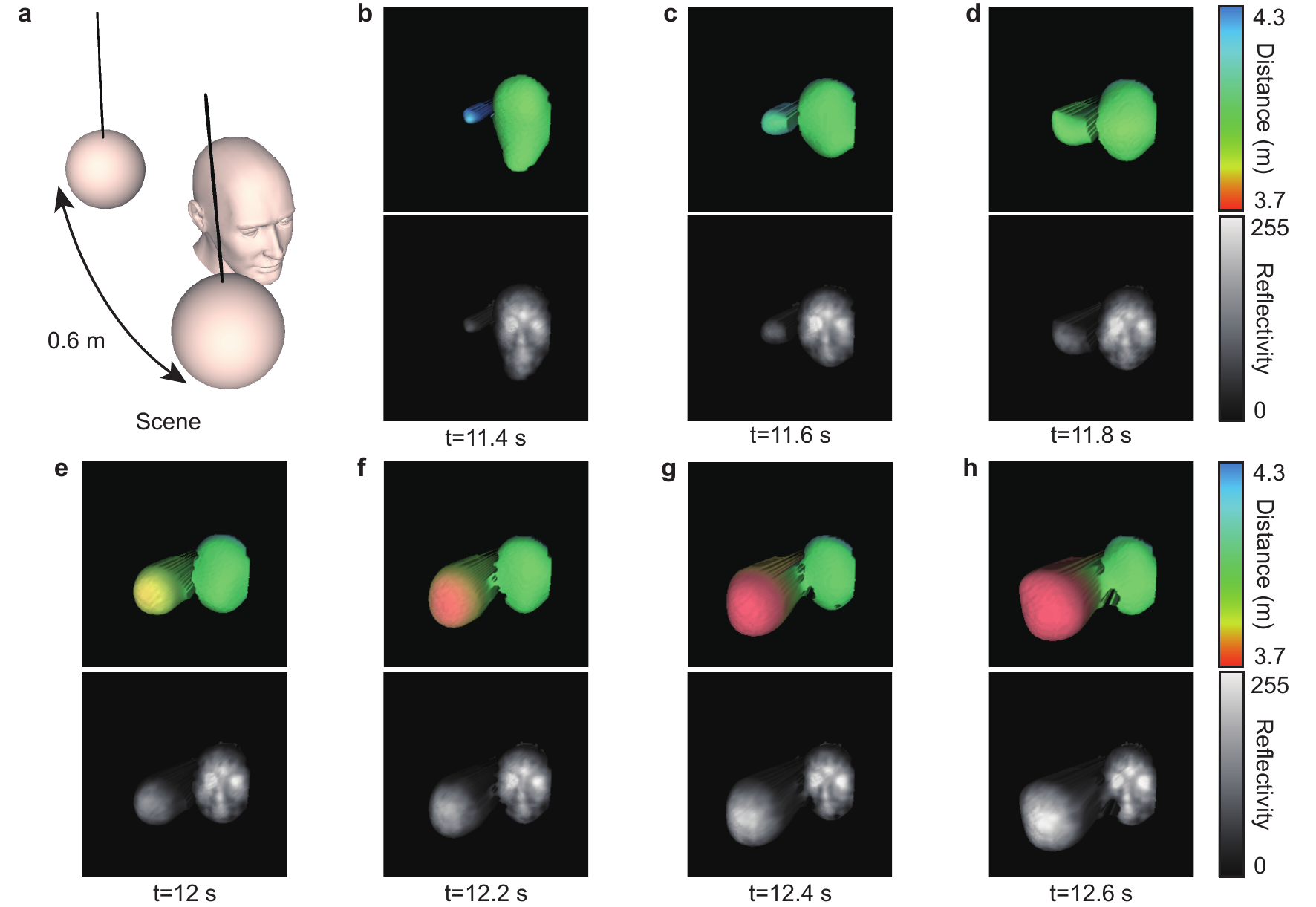}
	\caption{\textbf{Real-time 3D video.} (a) Illustration of the scene containing a static mannequin head and a swinging ball. (b)-(h) Sample of consecutive depth and reflectivity frames reconstructed at $\sim 5\,{\rm Hz}$ frame-rate in real-time for a transverse resolution of $64\times64$ pixels.}
	\label{Figure6}
\end{figure*}

In this work we employ an alternative scheme, known as evolutionary compressive sensing \cite{Radwell2014,Edgar2015}, which aims to reconstruct the image with significant less time than conventional compressive sensing by performing a linear iteration. In short, the evolutionary compressive sensing scheme chooses a subset of the Hadamard basis to display, by selecting the patterns with the most significant intensities measured in the previous frame, in addition to a fraction of randomly selected patterns that were not displayed. In this experiment, a scene consisting of a static polystyrene mannequin head and a polystyrene white ball ($140\,{\rm mm}$ diameter) swinging along the line of sight with a period of $\sim3\,{\rm s}$ (Fig.\,\ref{Figure6}a). The scene was located at a distance of $\sim 4\,{\rm m}$ from the imaging system. Two laser pulses were used per illumination pattern. With the approach described above we obtained continuous real-time 3D video with a frame-rate of $5\,{\rm Hz}$ using $600$ patterns (including their inverse) from the available $64\times64$ Hadamard set, equivalent to a compression ratio of $\simeq 7\%$. The experimental parameters for this result were chosen to balance the inherent trade-off between frame-rate and image quality. Figure\,\ref{Figure6}b-h show a sample of consecutive frames from the 3D video. The result shows an identifiable 3D image of the mannequin head and ball, in addition to the real-time motion of the ball. Importantly however, 3D reconsctruction can be performed using fewer patterns to achieve higher frame-rates if required, for instance using 256 patterns provides $12\,{\rm Hz}$ video.

\section*{Discussion}
We have demonstrated that our single-pixel 3D imaging system is capable of reconstructing a scene with millimetric ranging accuracy using modest hardware. Additionally, we obtained real-time video rates by taking advantage of a modified compressive sensing scheme that does not rely on lengthy post-processing.

The performance of the system in this work was mainly limited by the repetition rate of the laser employed, $7.4\,{\rm kHz}$. Using a laser with a repetition rate greater than or equal to the DMD modulation rate, could enable faster 3D video rates by a factor of three and/or increase reconstruction accuracy by increased averaging.
 
Furthermore, the broad operational spectrum ($400\,{\rm nm}-2500\,{\rm nm}$) of the DMD could enable the system to be extended to the non-visible wavelength, such as the infrared (IR), using modified source and detection optics. The use of DMDs in the IR have already been demonstrated in microscopy \cite{Radwell2014} and real-time video cameras \cite{Edgar2015}. The potential application of 3D imaging in the infrared could provide enhanced visibility at long-range, due to reduced atmospheric scattering \cite{Bucholtz1995Rayleigh}. 

\section*{Methods}
The following components were used in the experimental setup (Fig.\,\ref{Figure1}): a pulsed laser (Teem Photonics SNG-03E-100, $532\,{\rm nm}$), a DMD (Texas Instruments Discovery 4100 DMD), a projection lens (Nikon ED, f=$180\,{\rm mm}$), a collection lens (customised fresnel condenser lens, f=$20\,{\rm mm}$), a Si biased photodiode (Thorlabs DET10A) and a high-speed USB digitizer (PicoScope 6407, 2.5\,GS${\rm s^{-1}}$ for 2 channels acquisition). There are several important points worth mentioning. (a) The modulation rate of the DMD can reach up to $22.7\,{\rm kHz}$, however, in this experiment the DMD is operated in slave-mode, meaning the modulation rate is determined by the repetition rate of the laser at $7.4\,{\rm kHz}$. (b) The active area of the photodiode is $0.8\,{\rm mm^2}$, used in conjuction with a $20\,{\rm mm}$ focal length fresnel lens system, giving a $2.6\,^\circ$ field-of-view, which matches that of the projection system. The depth estimation includes Gaussian smoothing, intensity calibration, cubic spline interpolation, and depth determination.

\section*{Acknowledgements}
We would like to thank Prof. Adrian Bowman and Dr. Liberty Vittert for providing the stereophotogrammetric 3D data of the polystyrene mannequin head. MJP acknowledges financial support from UK Quantum Technology Hub in Quantum Enhanced Imaging (Grant No. EP/M01326X/1) the Wolfson foundation and the Royal Society. MS acknowledges the support from National Natural Foundation of China (Grant No. 61307021) and China Scholarship Council (Grant No. 201306025016).

\section*{Author contributions}
MS, RL and MJP conceived the concept of the experiment. MS, MPE and GMG. designed and performed the experiments. MS, MPE and MJP designed the reconstruction algorithm. NR developed the evolutionary compressed sensing algorithm.  MS, MPE, BS and MJP analysed the results. MS, MPE and MJP wrote the manuscript and other authors provided editorial input.

\section*{Competing financial interests}
The authors declare no competing financial interests.

\end{document}